\documentclass[10pt]{iopart}
\usepackage{iopams}
\usepackage{epsf}
\begin{document}

\title{Toward the search for gravitational waves from inspiraling compact binaries
in TAMA300 data during 2003: the data quality and stability}

\author{Hirotaka Takahashi\dag \ddag 
\footnote[3]{E-mail:hirotaka@vega.ess.sci.osaka-u.ac.jp},
Hideyuki Tagoshi\dag\ and the TAMA Collaboration}

\address{\dag\ Department of Earth and Space Science , 
Graduate School of Science, Osaka University, 
Toyonaka, Osaka 560--0043, Japan}

\address{\ddag\ Graduate School of Science and Technology, 
Niigata University, Niigata, Niigata 950-2181, Japan}

\begin{abstract}

We present the preliminary results of the analysis 
to search for inspiraling compact binaries 
using TAMA300 DT8 data which was taken during 2003. 
We compare the quality and the stability of the data with 
that taken during DT6 in 2001. 
We find that the DT8 data has better quality and stability 
than the DT6 data. 
\end{abstract}

\pacs{95.85.Sz, 04.80.Nn, 07.05Kf, 95.55Ym}




\vskip 0.5cm

TAMA300 \cite{tama} is an interferometric 
gravitational wave  detector with $300$ m baseline length located at 
Mitaka campus of the National Astronomical Observatory of Japan in Tokyo 
$(35.68 ^{\circ}N,139.54 ^{\circ}E)$.
TAMA300 began to operate in 1999. It had performed 
observations for 8 times by July 2003. 
In particular, during the period from 1 August to 20 September 2001, TAMA300 
performed an observation, which is called Data Taking 6 (DT6), 
and about 1039 hours of data were taken. 
The best sensitivity was about
$5\times10^{-21}/\sqrt{\rm Hz}$ around $800$Hz. 
More recently, 
a longer observation was performed by TAMA300 
during 14 February to 15 April. 2003, 
and about 1163 hours of data were taken. 
This observation is called Data Taking 8 (DT8).
The best sensitivity was about
$3\times10^{-21}/\sqrt{\rm Hz}$ around $1$kHz. 

We consider gravitational waves from inspiraling compact binaries, consisting of neutron 
stars or black holes. 
Since their wave forms are well known by post-Newtonian approximation 
of general relativity, we can use the matched filtering. 
In this paper, we report some results applied to DT6 and DT8 data,
and compare the stability of both observations. 
The data used for the analysis are that taken during 
1 August to 20 September, 2001 (DT6), 
and during 14 February to 15 April, 2003 (DT8). 
The length of data is 1039 hours and 1163 hours for DT6 and DT8 respectively. 
These are the data after removing any unlocked period, 
the period when some adjustments were made, and the locked period continued 
only less than 10 minutes. 

We assume that the time sequential data of the detector output $s(t)$ 
consists of a signal plus noise $n(t)$. 
We denote the one-sided power spectrum density of noise by $S_n(f)$. 

The gravitational wave strain amplitude 
$h(t)$ is calculated by combining two independent modes
of the gravitational wave and antenna pattern of interferometer as
\begin{equation}
h(t) = \mathcal{A} [h_c(t-t_c)\cos\phi_0 + h_s(t-t_c)\sin\phi_0],
\end{equation}
where $h_c(t)$ and $h_s(t)$ are two independent templates whose phase 
are different to $\pi/2$ from each other and $\mathcal{A}$ is amplitude.   
The parameters which characterize the templates are 
the coalescence time 
$t_c$, the chirp mass $\mathcal{M} (\equiv M\eta^{3/5})$ 
$(M =m_1+m_2)$, non-dimensional reduced mass 
$\eta (\equiv m_1 m_2 /M^2)$ and the phase $\phi_0$. 
The details of the wave form can be found in \cite{blanchet}. 
We use the restricted post-Newtonian wave forms 
in which the phase evolution is calculated to 2.5 
post-Newtonian order, 
but the amplitude evolution contains only the lowest 
Newtonian quadrupole contribution. 
The effects of spin angular momentum are not taken into account here. 

The two wave forms, $h_c$ and $h_s$, are transformed into the Fourier domain by 
the stationary phase approximation. 
We use templates for matched filtering defined in the Fourier domain as 
$\tilde{h}(f)=\tilde{h}_c(f)\cos\phi_0+\tilde{h}_s(f)\sin\phi_0$, 
where $\tilde{h}(f)$ denotes the Fourier transform of $h(t)$.
Here, $\tilde{h}_c$ and $\tilde{h}_s$ are normalized as
$(h_c|h_c)=$ $(h_s|h_s)=1$, where 
\begin{eqnarray}
(a|b)\equiv 2\int^{\infty}_{-\infty}df
{\tilde{a}(f)\tilde{b}^*(f)\over S_n(|f|)},
\end{eqnarray}
and * means the operation to take the complex conjugate.
The filtered output is then defined by 
\begin{equation}
\rho (t_c,M,\eta,\phi_0) \equiv (s|h). \label{eqn:rho} 
\end{equation}
In the matched filtering, we look for the maximum of $\rho$ 
over the parameter $t_c$, $M$, $\eta$, and $\phi_0$. 
In Eq. (\ref{eqn:rho}), we can take the maximization over $\phi_0$ analytically
which gives
\begin{equation}
\rho(t_c,m_1,m_2)
= \sqrt{(s|h_c) ^2 +(s|h_s) ^2}, \label{eqn:rho2} 
\end{equation}
We see that 
$\rho^2$ has an expectation value $2$ in the presence of only Gaussian 
noise. Thus, we can define the normalized signal-to-noise ratio by $SNR=\rho/\sqrt{2}$. 

In the TAMA300 analysis, 
we have found that the noise contained a large amount of 
non-stationary and non-Gaussian noise \cite{tama1}.
In order to remove the influence of such noise,
we introduce a $\chi^2$ test \cite{Allen}.
First we divide each template into $n$ mutually independent pieces
in the frequency domain,
chosen so that the expected contribution to $\rho$ from 
each frequency band is approximately equal, as 
\begin{equation}
\tilde{h}_{c,s}(f) = \tilde{h}_{c,s}^{(1)}(f)+ \tilde{h}_{c,s}^{(2)}(f)
+\cdots + \tilde{h}_{c,s}^{(n)}(f). 
\end{equation}
The $\chi^2$ is defined by
\begin{equation}
\chi^2 = n \sum_{i=1}^{n} \Big[{(z_{(c)}^{(i)}-\bar{z}_{(c)}^{(i)})^2
+(z_{(s)}^{(i)}-\bar{z}_{(s)}^{(i)})^2}\Big] . \label{eqn:chi}
\end{equation}
where
\begin{eqnarray}
z_{(c,s)}^{(i)}=(\tilde{s}|\tilde{h}_{c,s}^{(i)}) ,\quad\quad
\bar{z}_{(c,s)}^{(i)}=
{1\over n}
(\tilde{s}|\tilde{h}_{c,s}),
\end{eqnarray}
This quantity must satisfy the $\chi^2$-statistics with
$2n-2$ degrees of freedom, 
as long as the data consists of Gaussian noise plus 
inspiraling signals. 
For convenience, we use the reduced $\chi$-square defined by 
$\chi^2/(2n-2)$. In this paper, we chose $n=16$
(hereafter, $\chi^2$ means the reduced $\chi$-square).

The value of $\chi^2$ is independent of the amplitude of the signal as long as 
the template and the signal have an identical wave form. However, in reality, 
since the template and the signal have different value of parameters because of
the discrete time step and discrete mass parameters we search,  
the value of $\chi^2$ becomes larger when the amplitude
of signal becomes larger. In such situation, if we reject events simply by 
the value of $\chi^2$, we may lose real events with large amplitude. 
In order to avoid this, we have been introducing a statistic, 
$\rho/\sqrt{\chi^2}$, in the TAMA300 data analysis \cite{tagoshi}. 
The event selection using $\rho/\sqrt{\chi^2}$ is more efficient to detect 
events with large $\rho$ (say, $\rho>20$) than the usual $\chi^2$ selection. 
Further, by checking the Galactic event detection efficiency, we have found that 
the detection efficiency increases about 20 \% by the $\rho/\sqrt{\chi^2}$ 
selection compared to the $\chi^2$ selection in the case if the threshold is $\chi^2<1.5$.

The mass parameters considered in this paper is 
$1.0 M_{\odot} \le m_1,m_2 \le 2.0 M_{\odot}$, 
which is a typical mass region of neutron stars. 
In this region, a discrete mass parameter space is determined so that 
the maximum loss of SNR due to discretization 
becomes less than $3\%$. 
The parametrization of mass parameters are done based on \cite{tanaka-tagoshi}.

The mass parameter space depends on the power spectrum of noise. 
Averaged power spectrum of noise for each continuously locked segment
was used to define the mass parameter space in each segments. 
This parameter space is not changed within the continuously locked segment. 
However, in order to take into account of the variation
of the noise power spectrum with time, 
we use a different mass parameter space for different locked segments. 
The mass parameter space turned out 
to contain about $200 \sim 1000$ templates for the DT6 data, and about
$200 \sim 800$ templates for the DT8 data (Fig.1). 
The typical value of the number of template is about
700 for DT6, and 600 for DT8.
The variation of the number of template is not due to the variation of 
absolute amplitude of the power spectrum, but 
due to the variation of the shape of the power spectrum. 
Thus we can see an aspect of variation of the shape of the power spectrum
from Fig.1. 
It is found that DT8 data are more stable than DT6 data with respect to 
the number of templates, which probably means the stability of
DT8 data with respect to the shape of the power spectrum. 

Next we examine the variation of total noise power. 
Here, we define 
\begin{equation}
p=\left[4\int_0^\infty{f^{-7/3}\over S_n(f)}df\right]^{-1/2}. \label{eqn:p}
\end{equation}
Since this value is related to the normalization constant of the templates,
this noise power is suited to examine the variation of the noise power spectrum
used in the matched filtering. The value of $p$ is evaluated 
for each data with length 1.1 minutes. Here, we only consider the locked data. 
In Fig. 2, we plot the number of occurrence as a function of the value $p$. 
The mean value and standard deviation are 
$3.2 \times 10^{-19}$, $1.5 \times 10^{-19}$ for DT6 and
$1.9 \times 10^{-19}$, $7.5 \times 10^{-20}$ for DT8 
respectively. 
It is found that the stability of DT8 data is better than 
that of DT6 data 
with respect to the amplitude of the noise power spectrum. 

Next we present results of matched filtering analysis
including preliminary results of DT8 analysis. 
After calculating Eq.(\ref{eqn:rho2}) 
for each mass parameter in the mass parameter space,  
we search for $t_c$ and masses which realize the maximum value of $\rho$
in each interval of the coalescence time 
with length $\Delta t_c=25.6$ msec. 
For those $t_c$ and masses, the value of the reduced $\chi$-square are calculated. 
In Fig. 3, we plot the number of events as a function of $\rho$ and $\sqrt{\chi^2}$. 
In these figure, only events with $\rho>7$ are plotted. 
We find from the distribution of $\rho$ that the DT8 result has longer tail
than DT6 data. Such events with large $\rho$ must be due to the
non-Gaussian noise since the value of $\sqrt{\chi^2}$ of them are also very large. 
The distribution of $\sqrt{\chi^2}$ is also more spread in DT8 case than 
in DT6 case. 
Thus, in terms of the distribution of $\rho$ and $\sqrt{\chi^2}$, 
the stability of DT8 does not seem to be better than DT6. 
However, most of events with large $\rho$ in DT8 have larger $\sqrt{\chi^2}$
than that of DT6. Thus, if we plot the number of events as a function
of $\rho/\sqrt{\chi^2}$, we have a shorter tail in DT8 than in DT6. 
This can be seen in Fig.4. 
If we determine threshold of $\rho/\sqrt{\chi^2}$ for a given false alarm rate, 
the threshold of DT8 becomes lower than that of DT6. 
This fact will help us to improve the detection efficiency of the Galactic events
in DT8 further. These facts also suggest that the DT8 data have much better quality 
than the DT6 data in the sense
that it is easier to discriminate the non-Gaussian noise events. 

After the observation of DT6, the configuration of TAMA300 has been changed very 
much. One of the major changes is the installation of the power recycling system. 
Thus, it will be important and interesting to 
investigate the origin of the different property of data 
between DT6 and DT8. We will work on such analysis in the future. 

Much more details of the analysis using the data of TAMA300 DT6 and DT8 will 
be discussed elsewhere \cite{tagoshi}. 

\vspace{0.5cm}
This work was supported in part by Grant-in-Aid for Scientific Research,
Nos.~14047214 and 12640269, of the Ministry of Education, 
Culture, Sports, Science, and Technology of Japan.

\begin{figure}[htpd]
\begin{center}
\epsfysize= 7cm 
\epsfbox{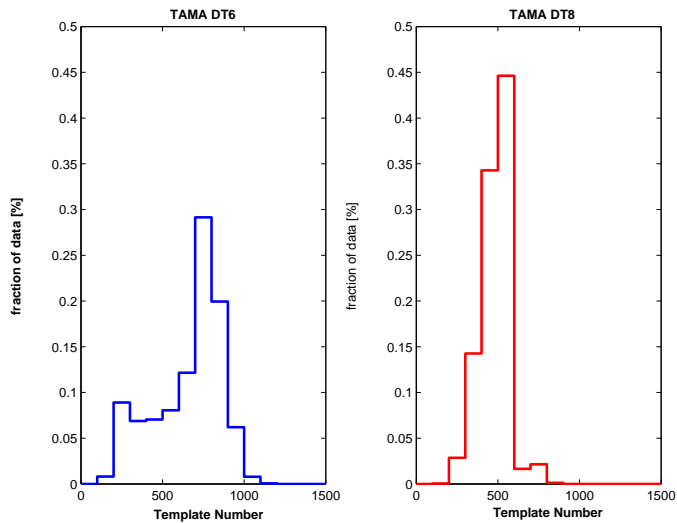}
\caption{The frequency distribution of the number of templates used in DT6 and DT8 analysis 
are plotted}
\end{center}
\end{figure}

\begin{figure}[htpd]
\begin{center}
\epsfysize= 7cm 
\epsfbox{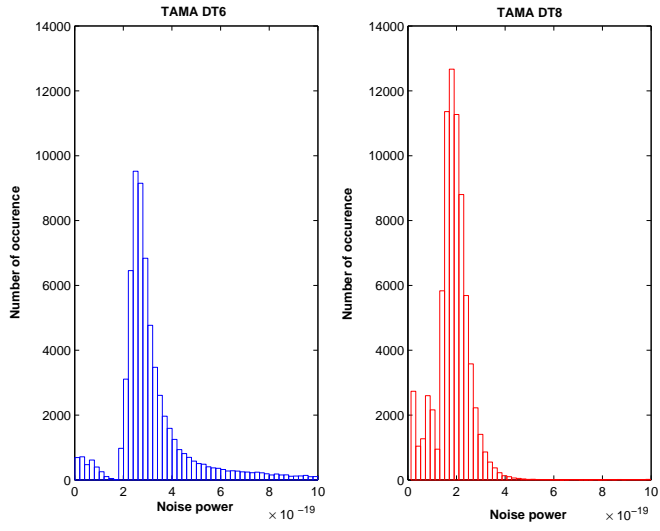}
\caption{Noise power $p$ for each locked data with 1.1 minutes.
The left and right panel shows the average of noise power for DT6 and DT8 
respectively.}
\label{fig:fig1}
\end{center}
\end{figure}

\begin{figure}[htpd]
\begin{center}
\epsfysize= 7cm 
\epsfbox{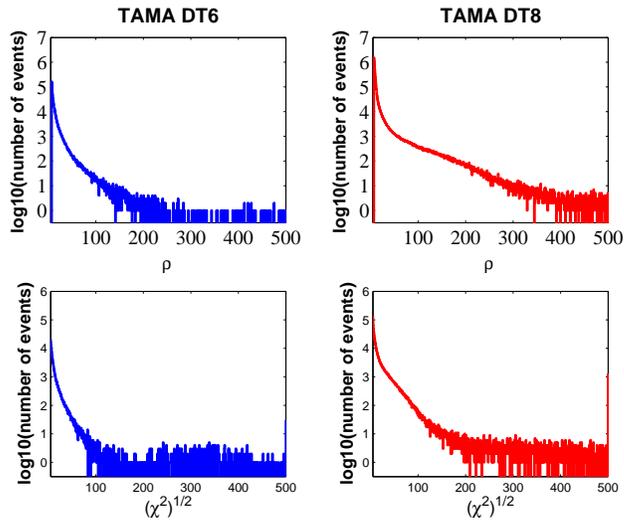}
\caption{The distribution of the number of events are plotted as a function of $\rho$
and $\sqrt{\chi^2}$. }
\end{center}
\end{figure}

\begin{figure}[htpd]
\epsfysize= 7cm 
\begin{center}
\epsfbox{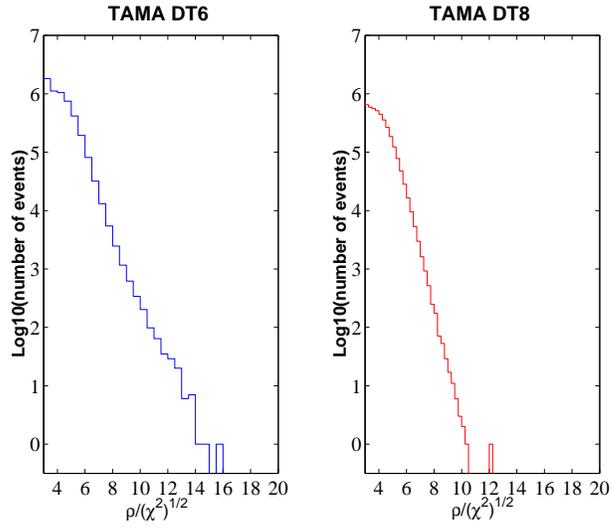}
\caption{The number of events as a function of $\rho/\sqrt{\chi^2}$. 
The left and right panel shows histograms for DT6 and DT8 respectively.}
\label{fig:fig2}
\end{center}
\end{figure}

\end{document}